\documentstyle[preprint,aps]{revtex}
\begin{document}
\draft
\preprint{SNUTP/96-113, hep-th/9611102}
\title{ Holstein-Primakoff Realizations on Coadjoint Orbits}
\author{Phillial Oh\cite{poh}}
\address{Department of Physics\\
Sung Kyun Kwan University\\
Suwon 440-746,  Korea}

\author{Chaiho Rim\cite{rim}}
\address{Department of Physics\\
Chonbuk National University\\
Chonju, 561-756,  Korea}

\date{\today}

\maketitle

\begin{abstract}
We derive the  Holstein-Primakoff oscillator 
realization  on the coadjoint orbits of the  
$SU(N+1)$ and $SU(1,N)$ group by treating
the coadjoint orbits as a constrained system and performing
the symplectic reduction.
By using the action-angle variables transformations,
we transform the original variables into Darboux
variables. The Holstein-Primakoff expressions emerge 
after quantization in a canonical manner 
with a suitable normal ordering. The corresponding
Dyson realizations are also obtained and 
some related issues are discussed.  

\end{abstract}

\newpage

\def\theequation{\arabic{section}.\arabic{equation}}
\section{Introduction}
\setcounter{equation}{0}

It is well known that the Holstein-Primakoff (HP) \cite{hols}
and Dyson \cite{dyso} realizations of
$su(2)$ algebra in terms of a single oscillator are very useful in describing the spin-density 
wave phenomena and many others in condensed matter physics \cite{kitt}
and nuclear physics \cite{klei}.
The HP realization also appears in the $q$-deformation \cite{ques} 
of the quantum algebras $su_q(2)$ and $su(1,1)_q$ \cite{bied} although 
$q$-deformation approach of 
of Jordan-Schwinger type \cite{bied1,sun} is more conventional.

Since the  HP and Dyson  
representations of $su(2)$ algebra can be interpreted as quantum
mechanical operators on $S^2$, which is the coadjoint orbit of
$SU(2)$ group, it is useful to consider them on the coadjoint
orbits of an arbitrary group in extending to higher group \cite{okub}.
So far, the generalization  was performed mostly to minimal 
$CP(N)$ orbits \cite{rand} or Grassmanian manifold \cite{mead} 
which was largely based on the coherent state method \cite{klau}.

In this letter, we discuss  general representations of HP 
and Dyson oscillator realizations for the  $su(N+1)$ and $su(1,N)$
algebras on the coadjoint orbits of $SU(N+1)$ and $SU(1,N)$
by treating the coadjoint orbits as a constrained classical system 
and by explicitly performing a symplectic reduction.
Compared with non-linear realization method on coset space
\cite{rand}, this approach can have some advantage of exploiting
the well-developed mathematical tool of symplectic reduction
\cite{arno} which in our case deals mainly with quadratic constraints.
The HP realization will emerge, 
if we transform the reduced system into canonical one by using
the action-angle variable and then quantize it 
in a standard manner with the normal ordering prescription.
Then, the Dyson realization will be obtained by shifting the 
square-root factor in HP realizations\cite{hech}. 
One of the merits of this coadjoint orbit approach is to 
provide a unified framework for finding  
explicit expressions for  HP and Dyson realization
in the compact and non-compact case. 
We will be mainly concerned with minimal and maximal orbits
of $SU(N+1)$ and $SU(1,N)$ to make the presentation simple.

We start by briefly explaining our notation.
Let us denote a column vector as a ket
$\vert Z>=(Z^{(1)},Z^{(2)},\cdots,Z^{(N+1)})^T$ and introduce
its bra $<\bar Z\vert= ( \bar Z_{(1)}, \bar Z_{(2)},\cdots , \bar Z_{(N+1)})$.
Then, $<\bar Z\vert Z>=\sum_{i=1}^{N+1}\bar Z_{(i)}Z^{(i)}
\equiv \bar Z_{(i)}M^{ij}Z_{(j)}$. The raising and lowering
are done with respect to the metric $M=\mbox{diag}(1,\epsilon,
\cdots,\epsilon)$.
$\epsilon=1(-1)$ for $SN(N+1)(SU(1,N))$. 
Let us express the element $g$ of $SU(N)$ and $SU(1,N)$ 
by $N+1$ kets $(\vert Z_1>, \vert Z_2>,\cdots, 
\vert Z_{N+1}>)$ with $ \vert Z_p>=(Z_p^{(1)},Z_p^{(2)},\cdots,Z_p^{(N+1)})^T$.
Then, $Mg^\dagger M$ is composed of $N+1$ bras
$<\bar Z^p\vert$'s such that $<\bar Z^p\vert=
( \bar Z^p_{(1)}, \bar Z^p_{(2)},\cdots , \bar Z^p_{(N+1)})$.
With the notation $< \bar Z_p\vert =< \bar Z^q\vert M_{qp}$, 
$Mg^\dagger Mg=I$ gives 
\begin{equation}
< \bar Z_p\vert Z_q>=M_{pq}, \quad \mbox{det}
(\vert Z_1>, \vert Z_2>,\cdots, \vert Z_{N+1}>) =1.
\label{cond}
\end{equation}

The isospin charges on the coadjoint orbits are defined by
\begin{equation}
Q^a=-2\mbox{Tr}(gxg^{-1}T^a). 
\end{equation}
where $x=i~\mbox{diag}(x_1,x_2,\cdots,x_{N+1})$ with
$\sum_{i=1}^{N+1}x_i=0$. The $x_i$'s are real and $T^a$'s are
the anti-hermitian generators of the group which satisfy the Lie
algebra with real structure constant $f_{ab}^{\ \ c}$:
$[T_a, T_b]=f_{ab}^{\ \ c}T_c$ and Tr$(T^aT^b)=-\frac{1}{2}\eta^{ab}$.
By making use of the second equation of the Eq.(\ref{cond}), 
$\vert Z_{N+1}>$ can be eliminated and subsequently we find that 
$Q^a$ can be expressed as \cite{oh955}
\begin{equation}
Q^a=-2i\sum_{p=1}^{N}J_p<\bar Z^p\vert T^a\vert Z_p>.
\label{isosp1}
\end{equation}
where $J_p=x_p-x_{N+1}=x_1+\cdots +2x_p+\cdots +x_{N}$.

Let us consider a classical system defined on the coadjoint orbit
of $SU(N)$ described by a Lagrangian
\begin{equation}
L=2\mbox{Tr}(xg^{-1}\dot g)-H(Q^a)=2
i\sum_{p=1}^{N+1}x_p <\bar Z^p\vert \frac{d}{dt}\vert Z_p>-H(Q^a).
\label{lagggg}
\end{equation}
By using the second equation of the Eq.(\ref{cond}) again, 
we find
\begin{equation}
L=2i\sum_{p=1}^{N}J_p<\bar Z^p\vert \frac{d}{dt}\vert Z_p>-H(Q^a).
\end{equation}
Note that there still exist the constraints 
$<\bar Z_p\vert Z_q>-M_{pq}=0\  (p,q=1,\cdots,N)$.

Using the symplectic structure of the above Lagrangian,
one can show that the isospin charges satisfy
the $su(N+1)$ and $su(1,N)$ algebras \cite{oh955}:
\begin{equation}
\{Q^a, Q^b\}=f^{ab}_{\ \ c}Q^c.
\end{equation}
HP realizations will be found if one finds a quantum mechanical
expression of the above isospin charges in terms of canonical 
variables and so it is essential to bring the Lagrangian
(\ref{lagggg}) into a canonical form. We will achieve this by
transforming the above system into action-angle variables.
In passing, we mention that action-angle variables approach
on the coadjoint orbits was also considered before \cite{alek} in the 
path integral quantization of the orbits in the compact case.  
 
\def\theequation{\arabic{section}.\arabic{equation}}
\section{Minimal Orbits }
\setcounter{equation}{0}
Let us first apply the above formalism to minimal orbits,
 $CP(N)$ and its non-compact counter part.
In this case, we have  $x=i~\mbox{diag}(J,-J/N,\cdots,-J/N)$ and
$J_1=J$, $J_2=\cdots=J_N=0$. 
In the compact case, $J$ is an integer for quantizable orbits.
For non-compact case, $J$ depends on the 
various types of representations of non-compact groups \cite{pere}.
With the notation 
$Z=(Z_0,Z_1,\cdots,Z_N)^T$ and introducing
$\bar Z= ( \bar Z_0, \bar Z_1,\cdots , 
\bar Z_N)$, we find that the Lagrangian can be written as 
\begin{equation}
L_Z=i J(\bar Z M \dot Z -\dot{\bar Z}MZ)-H(Q^a)
\label{lag1}
\end{equation}
with the constraint $\bar ZMZ=1$. Note that the notation in the above
equation denotes the conventional matrix product rather that the
abstract bracket inner product. In addition, the component is relabeled
from $0$ to $N$ instead of $1$ to $N+1$. We mention that
the above Lagrangian in the compact case was used in describing
the internal degrees of freedom of non-Abelian Chern-Simons
particles \cite{oh}.

It is well known that
the constraint can be solved explicitly in terms of the projective
coordinates defined by $\xi_i=Z_i/Z_0(Z_0\neq 0, i=1,2,\cdots,N)$
with a real gauge condition: \cite{alek,oh,ito}
\begin{equation}
\chi=\frac{1}{2}( Z^*_0 -Z_0) = 0\label{gauge1}.
\end{equation}
Then, the solution to the constraint $\bar ZMZ=1$ is given by
\begin{equation}
{\bar Z}_0=Z_0=\frac{1}{\sqrt{1+\epsilon\vert \xi\vert^2}},\ \ \ \ \vert\xi\vert^2
=\sum_i\vert\xi_i\vert^2,
\label{sol1}
\end{equation}
By substituting $Z_I=(Z_0,Z_0\xi_i)$ and Eq. (\ref{sol1}) into  
Eq. (\ref{lag1}), we obtain the following reduced Lagrangian: 
\begin{equation}
L_\xi=i J\epsilon \frac{{\bar \xi}\dot \xi-\dot{\bar \xi}\xi}
{1+\epsilon\mid\xi\mid^2}- H(Q^a).
\label{xilag}
\end{equation}
The isospin charges of Eq. (\ref{isosp1}) becomes 
\begin{equation}
Q^a=-i\frac{2J}{1+\epsilon\vert\xi\vert^2}
\left( T^a_{00}+T^a_{0i}\xi_i+\epsilon T^a_{i0}\bar\xi_i
+\epsilon T^a_{ij}\bar\xi_i\xi_j\right).
\label{defi}
\end{equation}

To make contact with HP representations, we make the following 
action-angle transformation of variables \cite{oh945}:
\begin{equation}
I_i=\frac{2\vert J\vert \vert\xi_i\vert^2}{1+\epsilon\vert\xi\vert^2}
\equiv \bar\alpha_i\alpha_i,
\label{darboux}
\end{equation}
and the angle variables are given by the phases of the $\alpha_i$'s.
Assuming a positive value for $J\epsilon$, we have 
the Lagrangian (\ref{xilag}) given by
\begin{equation}
L_\alpha=\frac{i}{2}({\bar \alpha}_i\dot \alpha_i-\dot{\bar \alpha_i}
\alpha_i) - H(Q^a).
\label{test}
\end{equation}
The Poisson bracket is defined in a canonical way
\begin{equation}
\{\alpha_i,\bar\alpha_j\}=i\delta_{ij}.
\label{canonical}
\end{equation}
Note that $J$ is negative in the
non-compact case. Otherwise, the role of $\alpha_i$
and $\bar\alpha_i$ would be interchanged in the canonical commutation
relation (\ref{canonical}). 
The  isospin functions (\ref{defi}) are expressed as follows:
\begin{eqnarray}
Q^a&=&-i\left[(2J-\epsilon\vert\alpha\vert^2)T^a_{00}
+\epsilon\bar\alpha_iT^a_{ij}\alpha_j\right.\nonumber\\
&+&\left.\epsilon\sqrt{2J-\epsilon\vert\alpha\vert^2}\bar\alpha_iT^a_{i0}+
\sqrt{2J-\epsilon\vert\alpha\vert^2}T^a_{0i}\alpha_i\right].
\label{isosp}
\end{eqnarray}

The quantum mechanical operator realizations are obtained
after quantizing the above operators by 
replacing the Poisson bracket (\ref{canonical}) with Dirac bracket 
$\alpha_i\longrightarrow a^\dagger_i,~\bar\alpha_i\longrightarrow
a_i$, and perform the normal ordering of the resulting
operators by putting the creation operators $a^\dagger$ 
to the left of annihilation operators $a$.
Following the above procedure, we get the following HP realization:
\begin{eqnarray}
\hat Q_{\mbox{hp}}^a&=&-i\left[(2J-\epsilon a^\dagger\cdot a)T^a_{00}
+\epsilon T^a_{ij}a^\dagger_j a _i\right.\nonumber\\
&+&\left.\epsilon\sqrt{2J-\epsilon a^\dagger\cdot a}T^a_{i0} a_i+
 a^\dagger_i T^a_{0i}
 \sqrt{2J-\epsilon  a^\dagger\cdot a}\right].
\label{isospin}
\end{eqnarray} 
If we shift the square root in front of $T^a_{i0} a_i$ to the back of
$a^\dagger_i T^a_{0i}$, we get the following  generalized 
Dyson realization:
\begin{eqnarray}
\hat Q_{\mbox{d}}^a&=&-i\left\{\epsilon
[ T^a_{i0}+ T^a_{ij}a^\dagger_j-
T^a_{00}a^\dagger_i-T^a_{0j}a^\dagger_ja^\dagger_i]a_i\right.\nonumber\\
&+&\left.2J T^a_{00}+2J T^a_{0i}a^\dagger_i\right\}.
\label{bargmannfock}
\end{eqnarray}
It is easy to check that the above realization satisfies
the algebras for both the compact and non-compact cases.
For the compact case with $\epsilon=1$, the above expression
was obtained as an holomorphic differential operator acting on
coherent state \cite{oh955}. Note that shifting the square root makes
$\hat Q^a$'s and $\hat Q^{a\dagger}$'s not manifestly conjugate to
each other in the Dyson case. To make the representation unitary, the
inner product should be defined \cite{hech} with respect to the
Liouville measure, while the Bargmann measure is used for the HP case. 

We note that a similar expression in the compact case
appeared in the study of the generalized spin system \cite{rand}.
Our result reduces to  it after a trivial rescaling of the variables
and choosing a specific representation. Our phase space is in the
canonical form (see the Eq. (\ref{canonical})) and the result holds for
arbitrary representation of the group. In addition, Eq. (\ref{isospin}) 
also covers the non-compact case. 

To put the above expressions into a more familiar form, 
we consider $SU(N+1)$ case. It is convenient to
use a representation in which  the the ladder operators, 
$E^a_{pq}\equiv E^{\alpha\beta}_{pq} ~(\alpha,\beta=1,
\cdots, N; p,q=0,1,\cdots N)$, are given as follows:
\begin{equation}
E^{0\alpha}_{pq}=-\delta^0_p\delta^\alpha_q,~~ 
E^{\alpha 0}_{pq}=-\delta^\alpha_p\delta^0_q,~~
E^{\alpha\beta}_{pq}=\delta^\alpha_p\delta^\beta_q~ (p\neq q) ,
\label{hp1}
\end{equation}
and the Cartan subalgebra is given by
$N$ diagonal matrices which are denoted by
$H^m\equiv T^{m^2+2m},~ m=1,2,\cdots N$
and expressed as follows:
\begin{equation}
H^m_{pq}=(\sum_{k=0}^{m-1}\delta_{pk}\delta_{qk}
-m\delta_{pm}\delta_{qm})/\sqrt{2m(m+1)}.
\label{hp2}
\end{equation} 
Then, we obtain the following HP realization from the 
Eq. (\ref{isospin}):
\begin{eqnarray}
\hat Q^{0i}&=&a_i^\dagger\sqrt{2J-a^\dagger\cdot a},
~~\hat Q^{i0}=\sqrt{2J-a^\dagger\cdot a}a_i=\hat Q^{0i\dagger}
\nonumber\\
\hat Q^{ij}&=&-a_j^\dagger a_i\label{hp4} ~~(i\neq j)\\
\hat Q^m&=&
\frac{1}{\sqrt{2m(m+1)}}\left((m+1)a^\dagger_m a_m
+\sum_{k=m+1}^Na^\dagger_k a_k -2J\right).\nonumber
\end{eqnarray}
The corresponding Dyson realizations of $SU(N+1)$ 
is given by \cite{rowe}:
\begin{eqnarray}
\hat Q^{0i}&=&a_i^\dagger(2J-a^\dagger\cdot a),
~~\hat Q^{i0}=a_i
\nonumber\\
\hat Q^{ij}&=&-a_j^\dagger a_i~~(i\neq j)\label{dy3}\\
\hat Q^m&=&
\frac{1}{\sqrt{2m(m+1)}}\left((m+1)a^\dagger_m a_m
+\sum_{k=m+1}^Na^\dagger_k a_k -2J\right).\nonumber
\end{eqnarray}

\def\theequation{\arabic{section}.\arabic{equation}}
\section{Maximal Orbits }
\setcounter{equation}{0}
Now, let us turn to the maximal orbits, flag manifold of the group.
 Here, in order to make the
presentation simple, we will restrict to the $SU(3)$ and $SU(1,2)$
case. Extension to higher group is straightforward. Let us choose the
element $x$ as $x=i \mbox{diag}(x_1,x_2,-(x_1+x_2))$.
Then, $J_1=2x_1+x_2, J_2=x_1+2x_2$. We require $x_1\neq x_2, 
J_1\neq 0, J_2\neq 0$.
Introduce again
$Z_i=(Z_{i0},Z_{i1},Z_{i2})^T ~(i=1,2)$ and $\bar Z_i= ( \bar Z_{i0}, 
\bar Z_{i1},\bar Z_{i2})$, we find
\begin{equation}
L=i\sum_{i=1,2}J_iM^{ii}
(\bar Z_iM \dot Z_i -\dot{\bar Z_i}MZ_i)-H(Q^a).
\label{lag2}
\end{equation}
The constraints are given by
\begin{equation}
\bar  Z_i M Z_j=M_{ij}
\label{constraint}
\end{equation}

To solve the constraints, we again choose the
real gauge conditions:
\begin{equation}
\bar Z_{10}=Z_{10}~(\neq 0),~~ \bar Z_{22}=Z_{22} ~(\neq 0).
\end{equation}
Defining the projective coordinates
$z_i=Z_{1i}/Z_{10},~ \xi_\alpha=Z_{2\alpha}/Z_{20}$
$(i=1,2;\alpha=0,1)$, the above constraints can be solved as
\begin{equation}
Z_{10}=\frac{1}{\sqrt{1+\epsilon\vert z\vert^2}},~~~
Z_{22}=\frac{1}{\sqrt{1+\epsilon\vert \xi_0\vert^2
+\vert \xi_1\vert^2}},
\label{sol2}
\end{equation}
and the remaining constraints become
$\xi_0=-\epsilon(\xi_1\bar z_1+\bar z_2)$.
To compare with the known results
of symplectic structure on the maximal orbit \cite{pick}, 
we redefine the variables
by $\bar\xi_1\rightarrow -z_3, \bar\xi_0\rightarrow -z_4$
with the remaining constraint  given by 
$z_4=\epsilon (z_2-z_1z_3)$.
Then the canonical one form of  the Lagrangian (\ref{lag2})
is given by $\theta=i(\partial-\bar\partial)W$, where
$W$ is given by
\begin{equation}
W=\log (1+\epsilon\vert z_1\vert^2+\epsilon\vert z_2\vert^2)^m
(1+\vert z_3\vert^2+\epsilon\vert z_2-z_1z_3\vert^2)^n
\label{symp}
\end{equation}
with $J_1=m$ and$J_2=-n$. In the compact case with $\epsilon=1$
and $m,n=$ integers, the above expression precisely 
reduces to the form given in Ref. \cite{pick}. For non-compact case,
they need not be integers.
From here on, we will use interchangeably
use the variables $z_1,z_2,z_3,z_4$ or
$z_1,z_2, w_0\equiv z_4, w_1\equiv z_3$.
With our new notation, the isospin function $Q^a$'s 
of the Eq.(\ref{isosp1}) becomes 
\begin{eqnarray}
Q^a&=&-i\frac{2m}{1+\epsilon\vert z\vert^2}
\left( T^a_{00}+T^a_{0i}z_i+\epsilon T^a_{i0}\bar z_i
+\epsilon T^a_{ij}\bar z_iz_j\right)\nonumber\\
&-&i\frac{2n}{1+\epsilon\vert w_0\vert^2
+\vert w_1\vert^2}
\left(-\epsilon T^a_{22}+\epsilon T^a_{2\alpha}\bar w_\alpha+
 T^a_{\alpha 2}\zeta_\alpha
-  T^a_{\alpha\beta}\bar w_\beta\zeta_\alpha\right)
\label{defii}
\end{eqnarray}
with $\zeta_\alpha=(w_0, \epsilon w_1)$.

Let us again consider the action-angle variable transformations
\begin{equation}
\alpha_i=\sqrt{2m}\frac{z_i}{\sqrt{1+
\epsilon\vert z\vert^2}},
~~\beta_\alpha=\sqrt{2n}\frac{w_\alpha}{\sqrt{1+
\epsilon\vert w_0\vert^2
+\vert w_1\vert^2}}
\label{darboux2}
\end{equation}
which renders the Lagrangian (\ref{lag2}) into a canonical
form 
\begin{equation}
L=\frac{i}{2}\left[ \epsilon({\bar \alpha}\dot \alpha-\dot{\bar \alpha}\alpha)
+\epsilon ({\bar \beta}m\dot \beta-\dot{\bar \beta}m\beta)
\right]- H(Q^a)
\label{can}
\end{equation}
with $m$ given by $m=\mbox{diag}(1,\epsilon)$.
We also have the isospin functions given by
\begin{eqnarray}
Q^a&=&-i\left[(2m-\epsilon\vert\alpha\vert^2)T^a_{00}
+\epsilon\bar\alpha_iT^a_{ij}\alpha_j+
\epsilon\sqrt{2m-\epsilon\vert\alpha\vert^2}\bar\alpha_iT^a_{i0}+
\sqrt{2m-\epsilon\vert\alpha\vert^2}T^a_{0i}\alpha_i
\right]\nonumber\\
&-&i\left[-\epsilon(2n-\epsilon\vert\gamma\vert^2)T^a_{22}
-\gamma_\alpha T^a_{\alpha\beta}\bar\beta_\beta
+\sqrt{2n-\epsilon\vert\gamma\vert^2}
\gamma_\alpha T^a_{\alpha 2}+
\epsilon\sqrt{2n-\epsilon\vert\gamma\vert^2}T^a_{2\alpha}
\bar\beta_\alpha\right].
\label{isosp2}
\end{eqnarray}
with $\gamma_\alpha=(\beta_0,\epsilon\beta_1)$.
The quantum mechanical operator realizations are obtained
after going through the same steps as in the
minimal case. 

To deal with the remaining constraint
$z_4=z_2-z_1z_3$, we will restrict to the compact case for
convenience.
One is tempted to substitute this constraint 
directly into the Eq.(\ref{isosp2}) and then quantize the system.
However, this would change the canonical structure 
of the Eq.(\ref{can}) in a very complicated manner.
Another way to carry out the analysis is to impose
the constraint on the quantum state. 
The constraint in terms of $\alpha_1,
\alpha_2, \alpha_3\equiv \beta_1, \alpha_4\equiv \beta_0$ is
given by
\begin{equation}
\Phi_{\mbox{hp}}=\alpha_4\sqrt{l_1}-\alpha_2\sqrt{l_2}+\alpha_1\alpha_3=0
\end{equation}
where $l_1=2m-\vert\alpha_1\vert^2-\vert\alpha_2\vert^2,~
l_2=2n-\vert\alpha_3\vert^2-\vert\alpha_4\vert^2$.
One can easily check that the constraints are second class.

Using the expression (\ref{isosp2}) and 
canonically quantizing the system, we obtain the following
HP realizations in the standard notation of the
generators $E$'s, the Eqs. (\ref{hp1}) and (\ref{hp2}):
\begin{eqnarray}
\hat Q^{1+i2}&=&a_1^\dagger\sqrt{\hat l_1}-a^\dagger_4a_3,~
\hat Q^{1-i2}=\sqrt{\hat l_1}a_1-a^\dagger_3a_4
\nonumber\\
\hat Q^3&=&a_1^\dagger a_1+\frac{1}{2}(a_2^\dagger a_2-
a_3^\dagger a_3+a_4^\dagger a_4)-m
\nonumber\\
\hat Q^{4+i5}&=&a_2^\dagger\sqrt{\hat l_1}+a^\dagger_4\sqrt{\hat l_2},~
\hat Q^{4-i5}=\sqrt{\hat l_1}a_2+\sqrt{\hat l_2}a_4\label{hpp}\\
\hat Q^{6+i7}&=&-a_2^\dagger a_1-a^\dagger_3\sqrt{\hat l_2},~
\hat Q^{6-i7}=-a_1^\dagger a_2-\sqrt{\hat l_2}a_3\nonumber\\
\hat Q^8&=&\frac{\sqrt{3}}{2}(a_2^\dagger a_2+
a_3^\dagger a_3+a_4^\dagger a_4)-\frac{1}{\sqrt{3}}m-\frac{2}{\sqrt{3}}n
\nonumber
\end{eqnarray}
where $\hat l_1=2m-a_1^\dagger a_1-a_2^\dagger a_2,
~ \hat l_2=2n-a_3^\dagger a_3-a_4^\dagger a_4$. Since the constraints
are second class, only half of the constraints is imposed on the physical state
\begin{equation}
\hat\Phi_{\mbox{hp}}\vert\mbox{phys}>
=(a^\dagger_4\sqrt{\hat l_1}-
a^\dagger_2\sqrt{\hat l_2}+a^\dagger_1a^\dagger_3)\vert\mbox{phys}>=0.
\label{coo}
\end{equation}
The physical states are labeled by $(m,n)$ and can be obtained
by  successive applications of $m$-times of
$a^\dagger_1, a^\dagger_2$ combined and $n$-times of
$a^\dagger_3, a^\dagger_4$ combined to the vacuum state.
The above condition (\ref{coo}) will give some restrictions on $m$ and $n$.
The result will determine irreducible representations of 
the $SU(3)$ group according to the Borel-Weil-Bott theorem\cite{kiri}. 
The detailed analysis on the relations between the
Eqs. (\ref{hpp}) and (\ref{coo}) and the irreducible representations
is not of concern here and will be reported elsewhere.
  
By shifting the square root, we again get the corresponding 
constrained Dyson realizations:
\begin{eqnarray}
\hat Q^{1+i2}&=&a_1^\dagger {\hat l_1}-a^\dagger_4a_3,~
\hat Q^{1-i2}=a_1-a^\dagger_3a_4
\nonumber\\
\hat Q^3&=&a_1^\dagger a_1+\frac{1}{2}(a_2^\dagger a_2-
a_3^\dagger a_3+a_4^\dagger a_4)-m
\nonumber\\
\hat Q^{4+i5}&=&a_2^\dagger{\hat l_1}+a^\dagger_4{\hat l_2},~
\hat Q^{4-i5}=a_2+a_4\label{diy}\\
\hat Q^{6+i7}&=&-a_2^\dagger a_1-a^\dagger_3{\hat l_2},~
\hat Q^{6-i7}=-a_1^\dagger a_2-a_3
\nonumber\\
\hat Q^8&=&\frac{\sqrt{3}}{2}(a_2^\dagger a_2+
a_3^\dagger a_3+a_4^\dagger a_4)-\frac{1}{\sqrt{3}}m-\frac{2}{\sqrt{3}}n
\nonumber
\end{eqnarray}
And we can infer that the constraint changes into
\begin{equation}
\hat\Phi_{\mbox{d}}\vert \mbox{phys}>=
(a^\dagger_4-a^\dagger_2+a^\dagger_1a^\dagger_3)\vert \mbox{phys}>=0.
\label{dyyyy}
\end{equation}
 
Let us compare the above formula with the other
Dyson realization which can be  obtained by the method of
geometric quantization in the holomorphic coherent state
approach \cite{oh1}:
\begin{eqnarray}
\hat Q^1&=&-\frac{1}{2}\left[(z_1^2-1)\frac{\partial}{\partial z_1}
+z_1z_2\frac{\partial}{\partial z_2}
+(z_2-z_1z_3)\frac{\partial}{\partial z_3}-2m z_1\right]
\nonumber\\
\hat Q^2&=&-\frac{i}{2}\left[-(z_1^2+1)\frac{\partial}{\partial z_1}
-z_1z_2\frac{\partial}{\partial z_2}
-(z_2-z_1z_3)\frac{\partial}{\partial z_3}+2m z_1\right]
\nonumber\\
\hat Q^3&=&z_1\frac{\partial}{\partial z_1}
+\frac{z_2}{2}\frac{\partial}{\partial z_2}
-\frac{z_3}{2}\frac{\partial}{\partial z_3}-m
\nonumber\\
\hat Q^4&=&-\frac{1}{2}\left[z_1z_2\frac{\partial}{\partial z_1}
+(z_2^2-1)\frac{\partial}{\partial z_2}
+z_3(z_2-z_1z_3)\frac{\partial}{\partial z_3}-2m z_2-2n(z_2-z_1z_3)\right]
\nonumber\\
\hat Q^5&=&-\frac{i}{2}\left[-z_1z_2\frac{\partial}{\partial z_1}
-(z_2^2+1)\frac{\partial}{\partial z_2}
-z_3(z_2-z_1z_3)\frac{\partial}{\partial z_3}+2m z_2+2n(z_2-z_1z_3)\right]
\label{dyy}\\
\hat Q^6&=&\frac{1}{2}\left[-z_2\frac{\partial}{\partial z_1}
-z_1\frac{\partial}{\partial z_2}
+(z_3^2-1)\frac{\partial}{\partial z_3}-2nz_3\right]
\nonumber\\
\hat Q^7&=&\frac{i}{2}\left[z_2\frac{\partial}{\partial z_1}
-z_1\frac{\partial}{\partial z_2}
-(z_3^2+1)\frac{\partial}{\partial z_3}+2nz_3\right]
\nonumber\\
\hat Q^8&=&\frac{\sqrt{3}}{2}(z_2\frac{\partial}{\partial z_2}
+z_3\frac{\partial}{\partial z_3})-\frac{1}{\sqrt{3}}m
-\frac{2}{\sqrt{3}}n.
\nonumber
\end{eqnarray}
We find that the expression (\ref{diy}) reduces to the above one
after naively using the Fock-Bargmann representation
$a_i\rightarrow \partial/\partial z_i, a_i^\dagger\rightarrow
z_i$ with the substitution $z_4=z_2-z_1z_3$ and acting
on the physical states annihilated by $a_4, \frac{\partial}
{\partial z_4}\vert\mbox{phys}>=0$. 
However, the relation between the two approach must be 
investigated further:
the Eqs. (\ref{diy}) and (\ref{dyyyy}) 
which correspond to the process of 
reduction after quantization, in general, does not give the same
result as the case of  quantization after reduction,
Eq. (\ref{dyy}).

\def\theequation{\arabic{section}.\arabic{equation}}
\section{Conclusion}
\setcounter{equation}{0}
We studied the  HP oscillator 
realization  on the coadjoint orbits of the  
$SU(N+1)$ and $SU(1,N)$ group by considering 
the symplectic reduction of these group 
and by using the action-angle variables transformations.
The HP expressions were obtained after canonical quantization
with a suitable normal ordering. In the minimal case,
the constraints can be solved explicitly but in the maximal case, 
some of the constraints were imposed directly on the physical states.
The corresponding Dyson realizations were also obtained.  

It would be  straightforward  to extend the above
formalism to other coadjoint orbits. Especially, it
would be interesting to apply it 
in studying the generalized spin system, ferromagnet or
antiferromagnet system on the flag manifold \cite{rand}
and the Hermitian symmetric space
\cite{ford,ohpark}.  
Finally, the $q$-deformation of the Eqs. (\ref{hpp}) and (\ref{diy})
poses another interesting problem. Details will appear elsewhere.

\acknowledgments
We would like to  thank J. Klauder and 
V. Manko for useful conversations.
This work is supported by the KOSEF
through the CTP at SNU and the project number
(96-1400-04-01-3, 96-0702-04-01-3),
and by the Ministry of Education through the
RIBS (BSRI/96-1419,96-2434).


\begin{references}
\bibitem{ito} A similar gauge fixing condition appears in 
C. Itoi and M. Kato, J. Phys. A: Math. Gen. {\bf 27}, 2915 (1994), 
which was used to obtain HP and Dyson realizations from the Schwinger 
representation by reducing the degrees of freedom. However, 
we note that this reduction does not work in general except for the 
$CP(N)$. For maximal orbits of the $SU(N)$, for example, 
the number of the oscillators is $N(N-1)/2$, which is no less than 
$N$ of the Schwinger representation when $N\geq 3$.
\end{references}
\end{document}